\documentclass[twocolumn,superscriptaddress,float,aps]{revtex4-1}
\usepackage{graphicx,amsfonts,amssymb,amsmath,hyperref,hypcap,enumerate}
\usepackage{color}
\usepackage{cleveref}

\newcommand{\bg}{\boldsymbol{g}}

\newcommand{\pp}{\boldsymbol{p}}
\newcommand{\qq}{\boldsymbol{q}}

\newcommand{\RR}{\boldsymbol{R}}
\newcommand{\rr}{\boldsymbol{r}}

\newcommand{\kk}{\boldsymbol{k}}
\newcommand{\dd}{\boldsymbol{d}}

\newcommand{\bb}{\boldsymbol{b}}

\newcommand{\GG}{\boldsymbol{G}}

\begin{document}
	\title{Three-Dimensional Topological Twistronics}
	
	\author{Fengcheng Wu}
	\author{Rui-Xing Zhang}
	\author{Sankar Das Sarma}
	\affiliation{Condensed Matter Theory Center and Joint Quantum Institute, Department of Physics, University of Maryland, College Park, Maryland 20742, USA}

	
	\begin{abstract}
		We introduce a theoretical framework for the new concept of three-dimensional (3D) twistronics by  developing a generalized Bloch band theory for 3D layered systems with a constant twist angle $\theta$ between successive layers. Our theory employs a nonsymmorphic symmetry that enables a precise definition of an effective out-of-plane crystal momentum, and also captures the in-plane moir\'e pattern formed between neighboring twisted layers. To demonstrate the novel topological physics that can be achieved through 3D twistronics, we present two examples. In the first example of chiral twisted graphite, Weyl nodes arise because of inversion-symmetry breaking, with  $\theta$-tuned transitions between type-I and type-II Weyl fermions, as well as  magic angles at which the in-plane velocity vanishes. In the second example of twisted Weyl semimetal, the twist in the lattice structure induces a chiral gauge field $\boldsymbol{\mathcal{A}}$ that has a vortex-antivortex lattice configuration. Line modes bound to the  vortex cores of the  $\boldsymbol{\mathcal{A}}$ field  give rise to 3D Weyl physics in the moir\'e scale. We also discuss possible experimental realizations of 3D twistronics.	
	\end{abstract}

	\maketitle

	{\it Introduction.---}
	Moir\'e superlattices formed in twisted bilayers lead to novel two-dimensional (2D) phenomena. In twisted bilayer graphene (TBG), there are magic twist angles, at which moir\'e bands become nearly flat due to vanishing Dirac velocity \cite{Bistritzer2011} and many-body interactions are effectively enhanced. TBG represents a prototypical system for 2D twistronics \cite{Carr2017}, where the twist angle serves as a new tuning parameter.  Given the greatly exciting 2D physics developing in TBG such as the discovery of superconducting and correlated insulating states \cite{Cao2018Super,Cao2018Magnetic, Dean2018tuning, lu2019superconductors,sharpe2019emergent,serlin2019intrinsic,Guinea2012,Senthil2018,Koshino2018,Kang2018,liu2018complete,Bernevig2018Topology}, it is natural to wonder whether the concept of twistronics can be generalized to 3D systems. 
	
\begin{figure}[t]
		\includegraphics[width=1\columnwidth]{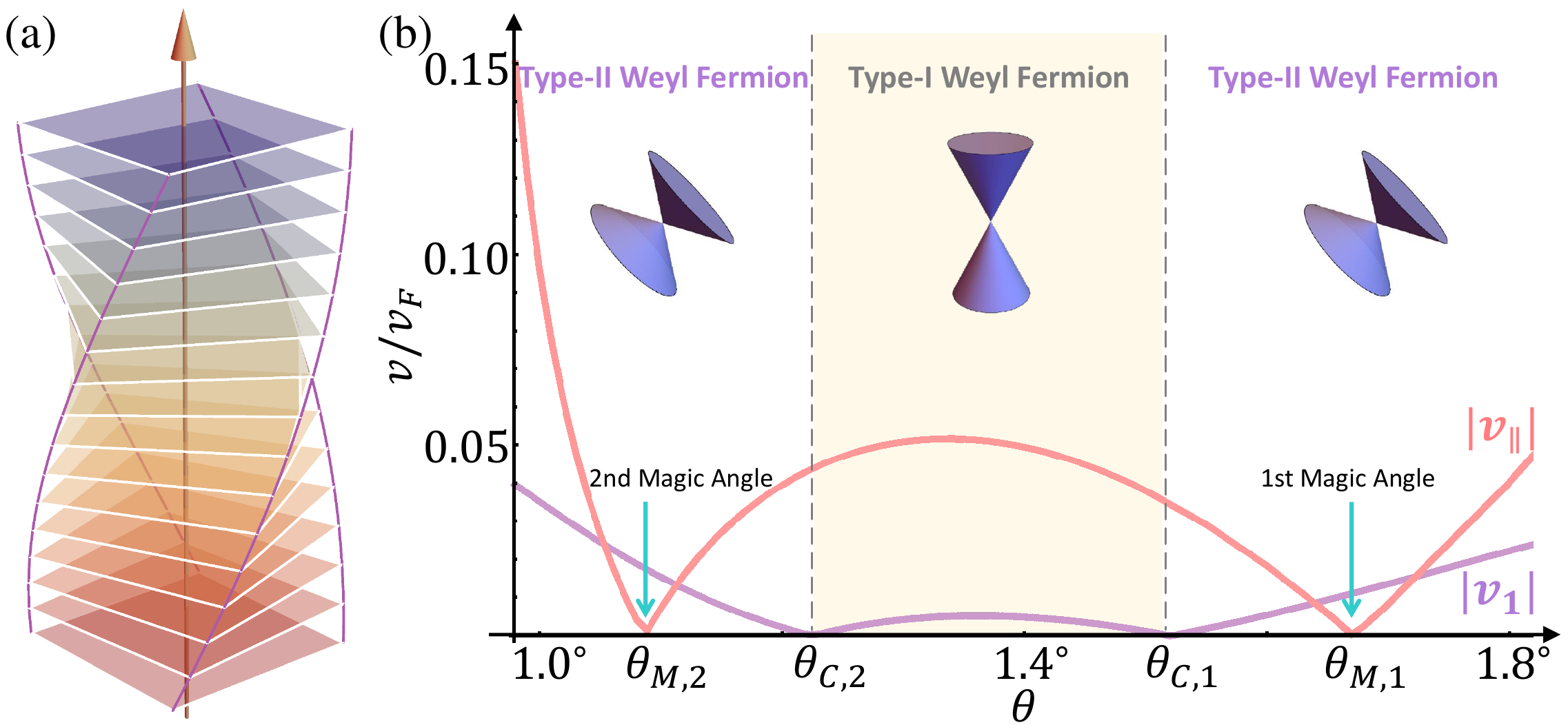}
		\caption{(a)Illustration of a 3D twisted structure with a constant twist angle $\theta$ between successive layers. (b) Summary of results on magic-angle Weyl fermions in chiral twisted graphite. The plot shows the in-plane velocity $v_{\parallel}$ and one of the out-of-plane velocities $v_1$ for the  Weyl fermion  at $\kk_{1/2}=(0,0,\pi/2)$. $v_{\parallel}$ vanishes at  magic angles $\theta_{M,1}$ and $\theta_{M,2}$. $v_1$ changes sign at $\theta_{C,1}$ and $\theta_{C,2}$, which mark transitions between type-I and type-II Weyl fermions. }
		\label{Fig1}
	\end{figure}

	In this Letter, we present a theoretical framework for 3D twistronics that can be realized in 3D layered systems with a constant twist angle $\theta$ between successive layers. This 3D chiral twisted structure [Fig.~\ref{Fig1}(a)] generally breaks the translational symmetry in all spatial directions and thus the conventional Bloch theorem cannot be applied. However, the structure has an exact nonsymmorphic symmetry, which consists of an in-plane $\theta$ rotation followed by an out-of-plane translation. We use this screw rotational symmetry to define a generalized Bloch's theorem, where the modified crystal momenta are well defined. Various 3D moir\'e physics can be explored by considering different 2D building blocks in our theoretical framework.
	
	We apply our theory to two  systems. In the first system of chiral twisted graphite with graphene as the 2D building block, Weyl fermions arise due to the inversion-symmetry breaking in the twisted structure. Both type-I and type-II \cite{soluyanov2015type} Weyl fermions can be realized depending on the value of $\theta$. Moreover, we find two  magic angles at which the in-plane Fermi velocity of the Weyl fermions vanishes, representing the realization of magic-angle Weyl physics for the first time. In the second system of twisted Weyl semimetal, we study effects of chiral twist in the lattice structure on Weyl fermions that already exist even without the twist.  The chiral twist induces a chiral gauge field $\boldsymbol{\mathcal{A}}$ that has a vortex-antivortex lattice configuration in the moir\'e pattern formed between adjacent twisted  layers. The vortex cores of the $\boldsymbol{\mathcal{A}}$  field bind line modes with position dependent chiralities, which generalizes the quasi-1D physics of a Weyl nanotube under torsion \cite{Franz2016} to 3D. The periodic array of the coupled  vortex line modes  gives rise to 3D Weyl fermions with moir\'e-scale modulations in the wave function. Therefore, the twist angle provides a new tuning knob to create and manipulate  Weyl fermions, and, more generally, topological phases in 3D.

	{\it Theory.---}
	We construct a generalized Bloch band theory for the chiral twisted structure shown in Fig.~\ref{Fig1}(a). The continuum Hamiltonian for this system is
	\begin{equation}
	\begin{aligned}
	H=\sum_{n} \int d^2 \rr \Big\{&\psi_n^\dagger(\rr) h_n (\kk_{\parallel}) \psi_n(\rr)
	\\
	&+[\psi_n^\dagger(\rr) T_n(\rr)\psi_{n+1}(\rr) + \text{H.c.}]
	\Big\},
	\end{aligned}
	\label{HH}
	\end{equation}
	where $n$ is the layer index, $\rr$ and $\kk_{\parallel}=-i\partial_{\rr}$ are respectively the 2D in-plane position and momentum operators, $\psi_n^\dagger(\rr)$ represents the field operator for low-energy states,
	$h_n(\kk_{\parallel})$ is the  in-plane  Hamiltonian for each 2D building block, and $T_n(\rr)$ is the interlayer tunneling. Here $\psi^\dagger$  can be a multicomponent spinor due to sublattices, orbitals, spins, etc. The layer dependence of $h_n$ and $T_n$ is determined by the twist relation:
	\begin{equation}
	h_n[\hat{R}(n\theta)\kk_{\parallel}]=h_0(\kk_{\parallel}),\,\,\, T_n[\hat{R}(n\theta)\rr]=T_0(\rr),
	\label{hnTn}
	\end{equation}
	where $\hat{R}$ is a rotation matrix. $T_0(\rr)$ has an in-plane moir\'e periodicity ($\propto 1/\theta$) when $\theta$ is small.
	
	The 3D twisted structure generally breaks translational symmetry in all spatial directions, making it appear hopeless for theoretical treatments. However, Eq.~(\ref{hnTn}) implies that the Hamiltonian $H$ is invariant under a nonsymmorphic operation, which rotates a layer by $\theta$ and then translates it along the out-of-plane $\hat{z}$ direction by the interlayer distance $d_z$. This nonsymmorphic symmetry suggests a generalized Bloch wave for the system:
	\begin{equation}
	\psi_{k_z}(\rr) = \frac{1}{\sqrt{N}} \sum_{n} e^{-i n k_z} \psi_n[\hat{R}(n\theta)\rr],
	\label{gBloch}
	\end{equation}
	where $N$ is the number of layers, and the good quantum number $k_z$ is an effective out-of-plane crystal momentum measured in units of $1/d_z$. This Bloch wave  is a superposition of electron states  on a spiral line around the screw-rotation axis, as illustrated by the purple lines in Fig.~\ref{Fig1}(a).  Under this generalized Bloch representation, the Hamiltonian $H$ becomes
	\begin{equation}
	\begin{aligned}
	H=&\sum_{k_z} \int d^2 \rr  \Big\{\psi_{k_z}^\dagger(\rr) h (\kk_{\parallel}) \psi_{k_z}(\rr)
	\\
	&+\Big[\psi_{k_z}^\dagger(\rr) e^{i k_z} T(\rr) \psi_{k_z}[\hat{R}(-\theta)\rr]+ \text{H.c.}\Big]
	\Big\},
	\end{aligned}
	\label{HBloch}
	\end{equation}
	where we use $h$ and $T$  as short-hand notations respectively for $h_0$ and $T_0$.
	It is worth noting that the appearance of $\hat{R}(-\theta)$ in Eq.~(\ref{HBloch}) signals the breaking of in-plane translation symmetries.  To proceed, we expand $T(\rr)$  by moir\'e harmonics:
	$T(\rr) = \sum_{\bg} T_{\bg} e^{i \bg \cdot \rr}$,
	where $\bg$ is a moir\'e reciprocal lattice vector.   $T(\rr) $ generates in-plane momentum scatterings specified by $\kk_{\parallel}'= \hat{R}(\theta) \kk_{\parallel}+\bg$. For low-energy physics, $|\kk_{\parallel}|$ is generally of the same order of magnitude as $|\bg|$, which is proportional to $\theta$. Thus,  $\hat{R}(\pm \theta)$ can be approximated by an identity matrix in the small $\theta$ limit, and the error is on the order $\theta |\bg| \ll |\bg|$. Under this approximation, $H$  acquires a moir\'e translational symmetry:
	\begin{equation}
	\begin{aligned}
	&H \approx \sum_{k_z} \int d^2 \rr  \psi_{k_z}^\dagger(\rr) [h(\kk_{\parallel})+\Delta(k_z,\rr)] \psi_{k_z}(\rr),\\
	&\Delta(k_z,\rr)=e^{i k_z} T(\rr) + e^{-i k_z} T^{\dagger}(\rr),
	\end{aligned}
	\label{Happrox}
	\end{equation}
	which gives rise to energy bands in the 3D momentum space spanned by $k_z$ and the in-plane moir\'e Brillouin zone. Eq.~(\ref{Happrox}) is our effective Hamiltonian for the 3D small-angle twisted system, which builds in exactly the nonsymmorphic symmetry and captures the moir\'e pattern formed in neighboring twisted layers.

	\begin{figure*}[t]
		\includegraphics[width=1.85\columnwidth]{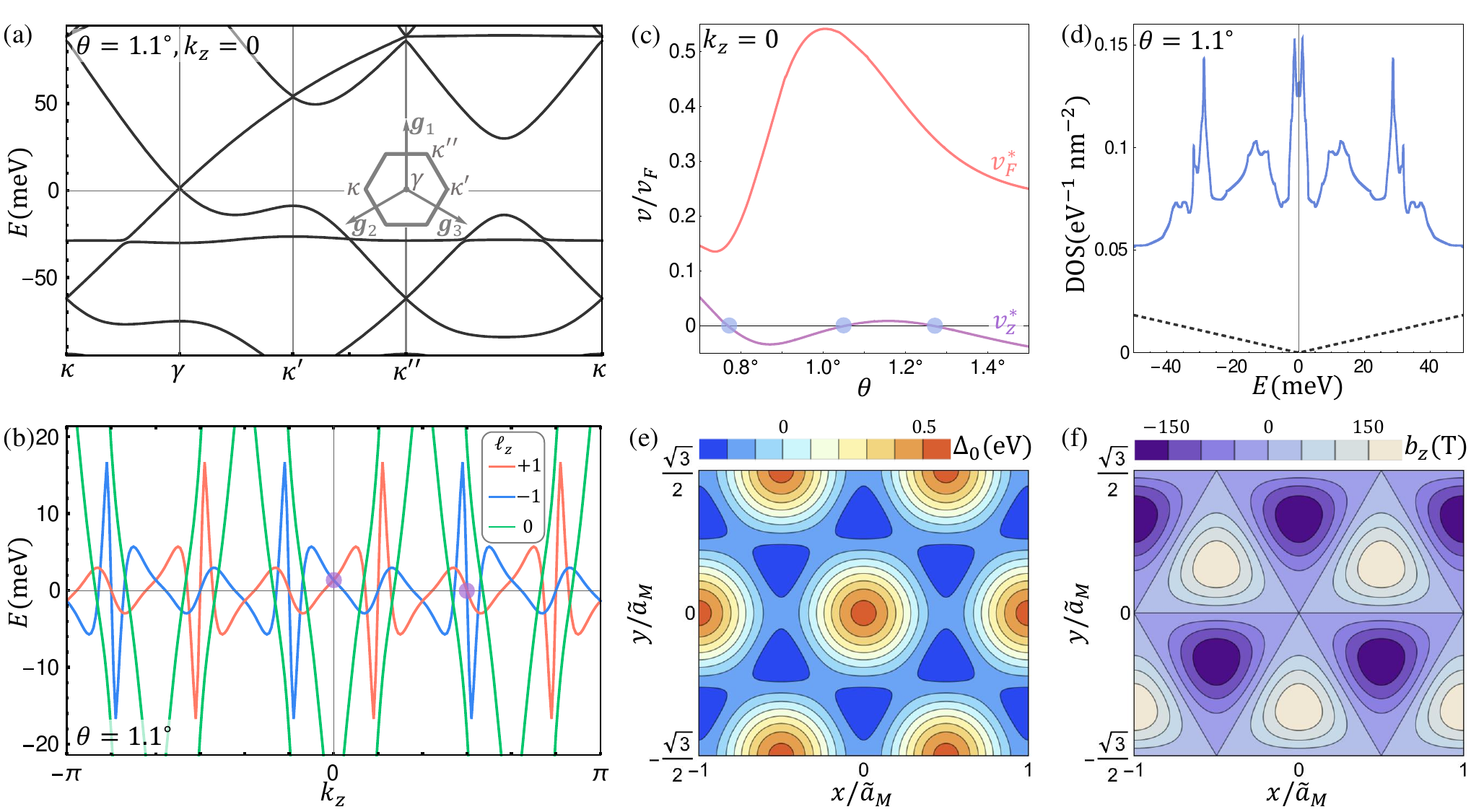}
		\caption{Results of chiral twisted graphite. (a) In-plane and (b) out-of-plane band structure. In (a) $k_z$ is 0. In (b), $\kk_{\parallel}$ is zero, the crossings between bands with different $\ell_z$ represent Weyl nodes, the two purple dots highlight nodes located at $k_z=0$ and $k_z=\pi/2$, and the spectrum  is periodic in $k_z$ with a period $2\pi/3$ due to a translation like symmetry $\mathcal{H}(\kk_{\parallel}, k_z+2\pi/3, \rr+a_M \hat{y})=\mathcal{H}(\kk_{\parallel}, k_z, \rr)$. (c) In-plane and out-of-plane velocities of the $\kk=0$ Weyl node as a function of $\theta$. (d) DOS per spin, valley, layer and area for the twisted graphite (blue line), and corresponding DOS for monolayer graphene (black dashed line). (e) 2D maps of the scalar moir\'e potential $\Delta_0$ and (f) pseudo magnetic field $b_z$ at $k_z=0$ and $\theta=1.1^\circ$.} 
		\label{Fig2}
	\end{figure*}
	
	\begin{figure*}[t]
		\includegraphics[width=1.85\columnwidth]{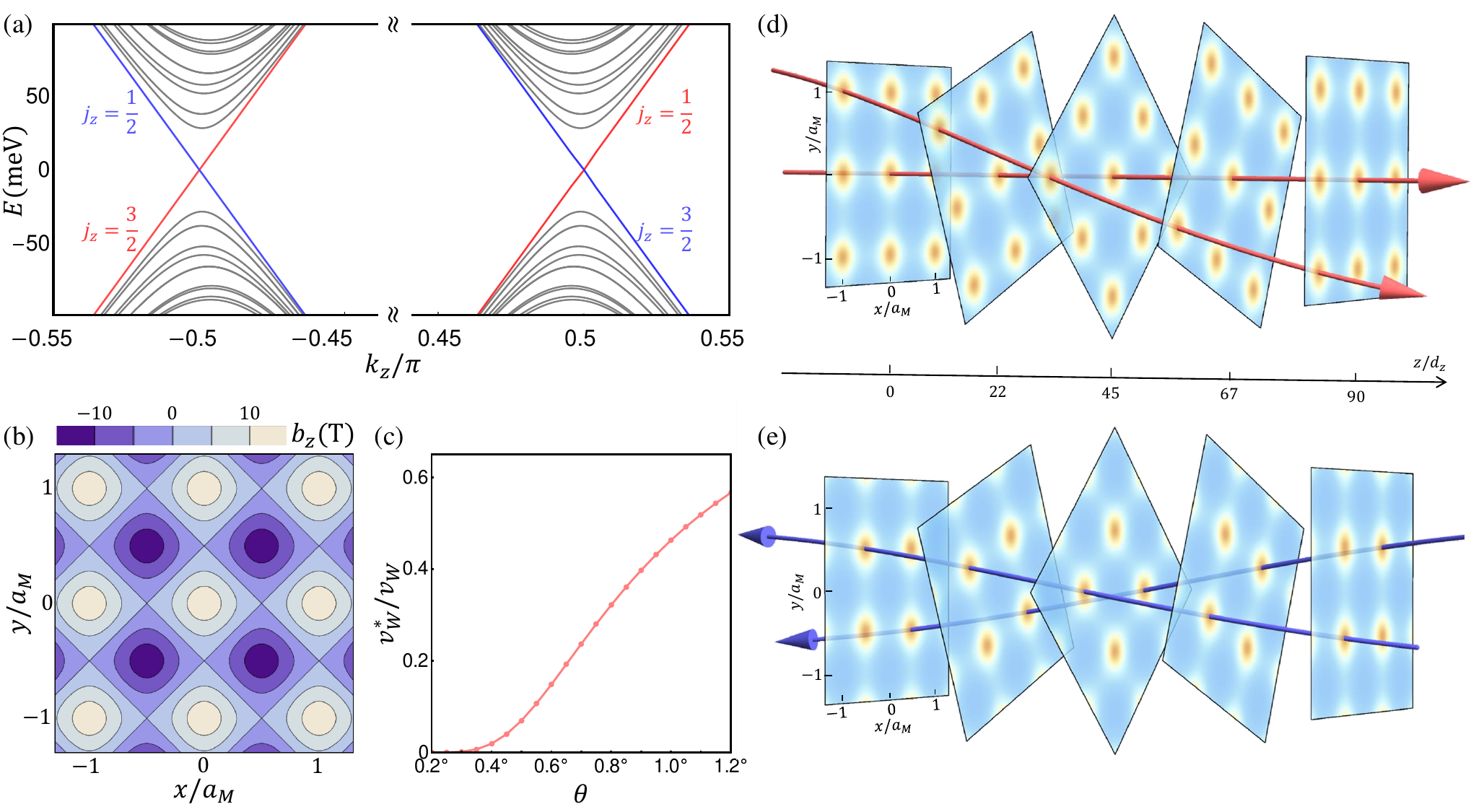}
		\caption{Results of twisted Weyl semimetal. (a) Out-of-plane band structure in the twisted Weyl semimetal. In-plane momentum is zero. Red and blue lines respectively mark right and left moving modes that cross Weyl nodes located at $k_z=\pm \pi/2$. $j_z$ specifies the angular momentum of each mode. Model parameters  are  $M_0=869$ meV, $M_1=10.36$ eV \AA$^2$, $t_{sp}=4$ meV, $v_W=3.74\times 10^5$ m/s, $a_0= 7.5$ \AA , and $Q_z=\pi/2$. (b) 2D maps of the pseudo magnetic field $b_z(k_z=\pi/2)$, which is  positive and negative, respectively, around integer $\boldsymbol{\mathcal{R}}$ and half-integer $\boldsymbol{\mathcal{R}}_{1/2}$ positions. $\theta=1^\circ$ in (a) and (b).  (c) The in-plane velocity $v_W^*$ of the Weyl node as a function of $\theta$. The out-of-plane velocity $v_z^*$ (not shown) barely changes with $\theta$. (d) Real-space probability density of the right moving modes in (a). The peaks of the density are at $\boldsymbol{\mathcal{R}}$  positions in each 2D layer, and wind around the screw rotation axis. (e) Similar as (d) but for the left moving modes in (a), and the peaks are located at  $\boldsymbol{\mathcal{R}}_{1/2}$  positions. }
		\label{Fig3}
	\end{figure*}

	{\it Chiral twisted graphite.---}
	We apply our theory to study electronic structure of chiral twisted graphite, which we construct by starting from an infinite number of graphene layers with $AAA...$ stacking, and then rotating the $n$th layer by  $n\theta$ around a common hexagon center. In each layer, low-energy electrons reside in $\pm K$ valleys, which are related by spinless time-reversal symmetry $\hat{\mathcal{T}}$ and can be studied separately as in TBG. We focus on $+K$ valley, with the in-plane $\kk \cdot \pp$ Hamiltonian   $h(\kk_{\parallel}) = \hbar v_F \kk_{\parallel} \cdot \boldsymbol{\sigma}$, where $v_F$ is the monolayer graphene Dirac velocity ($\sim 10^6 $m/s) and $\boldsymbol{\sigma}$ is the sublattice Pauli matrix. The interlayer tunneling $T(\rr)$ is \cite{Bistritzer2011,TBG2007}
	\begin{equation}
	T(\rr)=\sum_{j=0,1,2} \begin{pmatrix}
	w_{AA} & w_{AB} e^{-i 2\pi j/3} \\
	w_{AB} e^{i 2\pi j/3} & w_{AA}
	\end{pmatrix} e^{i \bg_{j+1} \cdot \rr}
	\label{Tgraphene}
	\end{equation}
	where $w_{AA}$ and $w_{AB}$ are  respectively intra-sublattice and inter-sublattice tunneling parameters, with $w_{AA} \approx 90$ meV and $w_{AB} \approx 117$ meV. $\bg_1$ is a moir\'e reciprocal lattice vector $(0,4\pi/3 a_M)$, and $a_M=a_0/\theta$, where $a_0$ is the monolayer graphene lattice constant. The other two vectors $\bg_{2,3}$ are related to $\bg_1$ by $\pm 2\pi/3$ rotations. We note that $a_M$ is the TBG moir\'e periodicity, but $T(\rr)$ in Eq.~(\ref{Tgraphene}) has a periodicity of $\tilde{a}_M= \sqrt{3} a_M$. 
	
	The $k_z$-dependent moir\'e potential $\Delta=e^{i k_z} T+e^{-i k_z} T^\dagger$ can be decomposed into $\Delta_0 \sigma_0+ \Delta_x \sigma_x+ \Delta_y \sigma_y $, where $\Delta_0$ is a scalar potential. From $\Delta$, we can define an effective gauge field $\boldsymbol{\mathcal{A}}=(\Delta_x, \Delta_y)/(e v_F)$ that couples to the Dirac Hamiltonian $h(\kk_{\parallel})$, and a corresponding pseudo magnetic field $b_z=\boldsymbol{\nabla}_{\rr} \times \boldsymbol{\mathcal{A}}$.
	2D maps of $\Delta_0$ and $b_z$ at $k_z=0$ are plotted in Fig.~\ref{Fig2}, which shows that $|b_z|$ can reach  $\sim$200 T for $\theta=1.1^\circ$.
	
	The effective Hamiltonian $\mathcal{H}=h(\kk_{\parallel})+\Delta$ respects  $\hat{C}_{3z}$  and $\hat{C}_{2z} \hat{\mathcal{T}}$ symmetries, where $\hat{C}_{n z}$ is the $n$-fold rotation around $\hat{z}$ axis. 
	We diagonalize $\mathcal{H}$ using a plane-wave expansion, and show the calculated band structures at $\theta=1.1^{\circ}$  in Fig.~\ref{Fig2}. Bands along $k_z$ axis can be characterized by the $\hat{C}_{3z}$ angular momentum $\ell_z \in \{0, \pm 1\}$.  As shown in Fig.~\ref{Fig2}(b),  crossings between two bands with different $\ell_z$  actually represent 3D Weyl nodes, which appear abundantly along $k_z$ axis. For example, the Weyl fermion at $\gamma$ point ($\kk=\boldsymbol{0}$) has an effective Hamiltonian  $\hbar(v_F^* \kk_{\parallel}\cdot \boldsymbol{\sigma}+v_z^* k_z \sigma_z)$, which is constrained  by both $\hat{C}_{3z}$  and $\hat{C}_{2z} \hat{\mathcal{T}}$ symmetries. The $\theta$ dependence of $(v_F^*,v_z^*)$ is shown in Fig.~\ref{Fig2}(c).  $v_F^*$ is reduced from the bare value $v_F$, but  remains finite for $\theta$ from $0.7^{\circ}$ to $2^{\circ}$. Remarkably, the sign of $v_z^*$ oscillates with $\theta$, and therefore, the chirality of the Weyl node at $\gamma$ point is twist angle dependent. 
	
	Another representative Weyl node is located at $\kk_{1/2}=(0,0,\pi/2)$. This Weyl node  is pinned to zero energy by a particle-hole like symmetry $\mathcal{H}(-\kk_{\parallel}, \pi-k_z, -\rr)=-\mathcal{H}(\kk_{\parallel}, k_z, \rr)$, and is described by $\hbar[v_{\parallel} \kk_{\parallel}\cdot \boldsymbol{\sigma}+v_1 q_z (\sigma_0+\sigma_z)/2+v_2 q_z (\sigma_0-\sigma_z)/2)]$, where $v_{1,2}$ are two independent parameters and $q_z=k_z-\pi/2$. For $\theta$ between $1^\circ$ and $1.8^\circ$, $v_2$ is always negative, but $v_1$ changes sign at $\theta_{C,1} \approx 1.52^\circ$ and $\theta_{C,2} \approx 1.22^\circ$, which are critical angles that mark  transitions between type-I and type-II Weyl fermions [Fig.~\ref{Fig1}(b)]. Moreover, the in-plane velocity $v_{\parallel}$ vanishes at both $\theta_{M,1} \approx 1.67^{\circ}$ and $\theta_{M,2} \approx 1.09^{\circ}$, which can be identified as two magic angles. Here the value of $\theta_{M,1}$ can also be estimated using an analytical perturbation theory, agreeing quantitatively with our direct band structure calculations; see Supplemental Material (SM)\cite{SM}.   The low-energy density of states (DOS) per layer in the twisted graphite near the magic angles is orders of magnitude larger than that in monolayer graphene [Fig.~\ref{Fig2}(d)], which should enhance interaction effects, leading to interaction driven quantum phase transitions.  
	
	We now compare our results with related works. Ref.~\onlinecite{Khalaf2019} studied multiple graphene layers with  twist angle $(-1)^{n} \theta$ that alternates with the layer index $n$. Their 3D structure preserves inversion symmetry, in contrast with our chiral twisted graphite structure. Ref.~\onlinecite{Cea2019} studied the same structure as ours but with a different method under the  coherent phase approximation. Our theory employs the exact nonsymmorphic symmetry, which allows us to precisely define the $k_z$ momentum and clearly demonstrate magic-angle Weyl physics in the twisted graphite. In Ref.~\onlinecite{Mora2019}, twisted trilayer graphene has been theoretically studied. An interesting question is how many layers are required  in practice to realize our predicted 3D physics, which we leave for future study.

	{\it Twisted Weyl semimetal.---}
	As another demonstration, we apply our theory to twisted Weyl semimetal with lattice structure also shown in Fig.~\ref{Fig1}(a). We start by introducing a minimal Weyl semimetal model on a simple cubic lattice (lattice constant $a_0$):
	$h_W = \hbar v_W \kk_{\parallel} \cdot \boldsymbol{\sigma} + M(\kk) \sigma_z$,
	where $M(\kk)=M_0(\cos k_z -\cos Q_z)- M_1 \kk_{\parallel}^2$ with parameters $M_{0,1}>0$ and $0<Q_z<\pi$.  Each site on the cubic lattice  accommodates two orbitals $|j_z=\frac{3}{2} \rangle$ and $|j_z=\frac{1}{2}\rangle$, which have different angular momentum $j_z$  and form the basis of the Pauli matrices $\boldsymbol{\sigma}$. The  Hamiltonian $h_W$ breaks time reversal symmetry and hosts  two Weyl nodes with opposite chiralities located respectively  at $\kk_{\pm}=(0,0,\pm Q_z)$. 
	
	We now consider a twisted cubic lattice in which the $n$th layer is rotated by $n\theta$ around $\hat{z}$ axis following Fig.~\ref{Fig1}(a). The effective Hamiltonian $\mathcal{H}_W$ for this twisted structure is given by
	\begin{equation}
	\begin{aligned}
	\mathcal{H}_W=h_W
	+ 2 {t_{sp}} \sin k_z \sum_{\bg}^{'} \begin{pmatrix}
	0 &  e^{-i (\phi_{\bg}+\bg\cdot \rr)} \\
	e^{i (\phi_{\bg}+\bg\cdot \rr)} & 0
	\end{pmatrix} 
	\end{aligned}
	\label{HWeyl}
	\end{equation}
	where  $t_{sp}$ characterizes the inter-orbital tunneling between neighboring layers. The summation over $\bg$ is restricted to the first shell of moir\'e reciprocal lattice vectors  ($|\bg|=2\pi/a_M$ with moir\'e period $a_M$ equal to $a_0/\theta$), and $\phi_{\bg}$ is the orientation angle of $\bg$. Here the interlayer tunneling matrix is derived using a two-center approximation, and  the spatial modulation of interlayer intra-orbital tunneling is neglected, which are thoroughly explained in the SM \cite{SM}. Compared with the twisted graphite Hamiltonian, $\mathcal{H}_W$ does not have a scalar moir\'e potential term. 
	
	We can extract an effective gauge field $\boldsymbol{\mathcal{A}}$ from $\mathcal{H}_W$, similar to the case of twisted graphite.
	The $\boldsymbol{\mathcal{A}}$ field has a vortex-antivortex lattice configuration in the moir\'e superlattices   \cite{SM}. The corresponding  pseudo magnetic field $b_z$ is given by
	\begin{equation}
	b_z= \frac{2 t_{sp}}{e v_W} \frac{2\pi}{a_M} \sin k_z \sum_{\bg}^{'}   e^{i \bg\cdot \rr},
	\end{equation}
	which is illustrated in Fig.~\ref{Fig3}(b). $|b_z|$ peaks at two distinct types of  positions in the moir\'e pattern, namely, integer positions $\boldsymbol{\mathcal{R}}=(n_x,n_y)a_M$ with $n_{x, y} \in \mathbb{Z}$  and half-integer positions $\boldsymbol{\mathcal{R}}_{1/2}=\boldsymbol{\mathcal{R}}+(\frac{1}{2},\frac{1}{2})a_M$, which are also locations of vortex cores of the $\boldsymbol{\mathcal{A}}$ field. We assume $t_{sp}/v_W>0$, and $b_z$ is positive and negative respectively at $\boldsymbol{\mathcal{R}}$ and $\boldsymbol{\mathcal{R}}_{1/2}$ for $0<k_z<\pi$. Being proportional to $\sin k_z$, $b_z$ represents a chiral magnetic field as it couples oppositely to the Weyl nodes with different topological charges. In the semiclassical picture, around $\boldsymbol{\mathcal{R}}$ ($\boldsymbol{\mathcal{R}}_{1/2}$) positions, this chiral magnetic field leads to chiral Landau levels  that propagate along positive (negative) $\hat{z}$ direction  for both  Weyl nodes at $\kk_{\pm}$. These chiral Landau levels bound to vortex cores of the $\boldsymbol{\mathcal{A}}$ field can be considered to be vortex line modes (VLMs).  The chirality of a VLM depends on its position in the moir\'e pattern. An important physical consequence is that  an out-of-plane electric field can drive a real-space pumping of electrons from $\boldsymbol{\mathcal{R}}$ to $\boldsymbol{\mathcal{R}}_{1/2}$ positions, or vice versa.  The 2D array of VLMs can further hybridize with each other, and we expect them to realize Weyl fermions with moir\'e-scale modulations in the wave functions.  
	
	To verify the above picture, we numerically diagonalize $\mathcal{H}_W$, and show the energy spectrum along $k_z$  in Fig.~\ref{Fig3}(a), where we find that the Weyl  nodes are robust against the twist. In Fig.~\ref{Fig3}(d) [Fig.~\ref{Fig3}(e)], we plot the real-space probability density of the right (left) moving modes highlighted in Fig.~\ref{Fig3}(a), which is found to be  primarily concentrated at $\boldsymbol{\mathcal{R}}$  ($\boldsymbol{\mathcal{R}}_{1/2}$) positions. This density profile is consistent with the above semiclassical analysis.  Because of the twisted structure, these  modes track spiral lines in real space. From  the semiclassical picture, the in-plane velocity $v_W^*$ of the 3D Weyl fermions is controlled by the coupling strength between neighboring VLMs, and therefore, by the moir\'e period $a_M$.
	Numerical results plotted in  Fig.~\ref{Fig3}(c)
	confirm that $v_W^*$ decreases with decreasing $\theta$ (equivalently, increasing $a_M$). For small enough $\theta$, $v_W^*$  nearly vanishes, showing that VLMs located at different positions are  essentially decoupled. Thus, the twist angle provides a new tuning knob to control the band structure of 3D Weyl materials.

	{\it Conclusion.---}
	In summary, we develop a general theoretical framework for 3D twistronics and apply the theory to chiral twisted graphite and twisted Weyl semimetal. Our theory can in principle be realized in van der Waals heterostructures constructed by stacking multiple twisted layers.  Moreover, chiral twisted van der Waals nanowires have recently been experimentally synthesized\cite{Sutter2019,Liu2019}, indicating that the chiral twisted structure we study can indeed appear in materials. Similar to 2D twisted bilayers, 3D twisted systems can provide new playgrounds for strongly correlated physics. As an example, superconducting instability should be enhanced in 3D twisted graphite near the magic angles due to the strong velocity suppression. In addition to solid state materials, our theory should be realizable in photonic and phononic systems, where 2D Dirac physics and 3D Weyl physics have been demonstrated\cite{rechtsman2013photonic,lu2015experimental,yang2018ideal,wang2019photonic,jia2019observation,xiao2015synthetic,yang2016acoustic,li2018weyl,dorrell2019van}. The flexibility of building metamaterials makes them promising platforms for our proposed 3D topological twistronics. Our predictions can be tested in experimental systems as disparate as suitably designed photonic structures and engineered solid state heterostructures as well as in layered systems made from graphite and manually stacked multilayer graphene.

	We thank Yang-Zhi Chou for discussions. 
	This work is supported by the Laboratory for Physical Sciences and Microsoft. R.-X. Z. is  supported by a JQI Postdoctoral Fellowship.

	\bibstyle{apsrev4-1} 
	\bibliography{refs}

\newpage
\clearpage
\setcounter{figure}{0}
\setcounter{equation}{0}
\renewcommand{\theequation}{S\arabic{equation}}
\renewcommand{\thefigure}{S\arabic{figure}}
\renewcommand{\thesection}{S\arabic{section}}

\begin{widetext}
	
	\begin{center}
		\textbf{Supplemental Material}
	\end{center}

	This Supplemental Material includes four sections. Sections~\ref{sec:1} and \ref{sec:2} provide additional discussion, respectively, on the generalized Bloch theory and on chiral twisted graphite.   Section~\ref{sec:3} is on a perturbation theory that provides a quantitative estimation of the first magic angle $\theta_{M,1}$ in chiral twisted graphite. Section~\ref{sec:4} presents a derivation of the interlayer tunneling matrix in twisted Weyl semimetal.

\section{Generalized Bloch Theory}
\label{sec:1}
In this section, we provide additional discussions on the generalized Bloch theory.

We note that the 1D version of our generalized Bloch that employs the nonsymmorphic symmetry has been considered previously in the literature, for example, Ref.~\onlinecite{Zheng1990}.

For the chiral twisted structure in Fig. 1 of the main text, we assume open boundary condition along $\hat{z}$ direction and an infinite number of layers. Under this assumption, the nonsymmorphic symmetry is exact, since each layer has a successive partner. The effective out-of-plane momentum $k_z$ is a good quantum number due to this nonsymmorphic symmetry and takes continuous values between $-\pi$ and $+\pi$. With this open boundary condition, the twist angle $\theta$ between adjacent layers does NOT have to be  rational in the form of $2 \pi p /q$, where $p$ and $q$ are integers.

If periodic boundary condition is imposed along $\hat{z}$ direction, the nonsymmorphic symmetry is exact only if $\theta$ takes rational values of $2 \pi p /q$ and $N \theta$ is an integer multiple of $2\pi$. In this case, $N$, the number of layers, can be finite, and $k_z$  takes discrete values between $-\pi$ and $+\pi$. Irrational $\theta/(2\pi)$ can be approximated by a rational number when periodic boundary condition is used.

In the derivation that leads to Eq. (5) of the main text, we approximate the rotation matrix $\hat{R}(\theta)$ by an identity matrix. Overall, this approximation is to neglect effects on the order of  $O(\theta^2)$. In real space, this is equivalent to neglect effects of ``moir\'e of moir\'e'' that appear on the length scale of $a_M/\theta$. We expect our theory to be accurate for system size below $a_M/\theta$.  For  very small twist angles (e.g. $0.1^{\circ}$), $a_M/\theta$ is much greater than $a_M$, which indicates that our theory should have a wide application range regarding the system size. 

In the main text, we only keep interlayer tunneling terms between adjacent layers. Tunnelings between layers that are further apart also respect the nonsymmorphic symmetry and can be treated using our theoretical framework but with some modifications. For example, the twist angle between next-nearest-neighbor layers is $2 \theta$, and the corresponding tunneling terms have a periodicity of $2 a_M$ instead of $a_M$, which can lead to further Brillouin zone folding effect in the momentum space. We expect that the topological physics that we study remains robust for weak remote out-of-plane tunneling effects.

\section{Hamiltonian and Symmetry of Chiral Twisted Graphite }
\label{sec:2}

We present a derivation of the Hamiltonian of chiral twisted graphite by starting from that of twisted bilayer graphene (TBG), and discuss the symmetries of the Hamiltonian and their implications on Weyl points. 

The moir\'e Hamiltonian for TBG in $+K$ valley is given by
\begin{equation}
\tilde{\mathcal{H}}_{\text{TBG}} = \begin{pmatrix}
\tilde{h}_0 (\kk_\parallel) & \tilde{T}(\rr)  \\
\tilde{T}^{\dagger}(\rr) & \tilde{h}_1 (\kk_\parallel)
\end{pmatrix},
\label{HTBG}
\end{equation} 
where $\tilde{h}_0 (\kk_\parallel)$ and $\tilde{h}_1 (\kk_\parallel)$ are Dirac Hamiltonians associated respectively with the bottom  and top  layers, $\tilde{h}_0 (\kk_\parallel)=\hbar v_F (\kk_\parallel - \boldsymbol{\kappa}_+)\cdot \boldsymbol{\sigma}$ and $\tilde{h}_1 (\kk_\parallel)=\hbar v_F e^{i \theta \sigma_z} [(\kk_\parallel - \boldsymbol{\kappa}_-)\cdot \boldsymbol{\sigma} ]$. Here $\boldsymbol{\kappa}_{\pm}$, equal to $[4\pi/(3 a_M)](-\sqrt{3}/2, \mp 1/2)$,  capture the layer-dependent momentum shift of the Dirac cones. $a_M$ is $a_0/\theta$, where $a_0$ is the lattice constant of mononlayer graphene. The interlayer tunneling is given by
\begin{equation}
\begin{aligned}
&\tilde{T}(\rr)  = T_1 + T_2 e^{- i \bb_+ \cdot \rr} + T_3 e^{- i \bb_- \cdot \rr}, \\
&T_j= \begin{pmatrix}
w_{AA} & w_{AB} e^{-i 2\pi (j-1)/3} \\
w_{AB} e^{i 2\pi (j-1)/3} & w_{AA}
\end{pmatrix},
\end{aligned}
\label{Trr}
\end{equation}
where $\bb_{\pm}$ represent TBG reciprocal lattice vectors given by $[4\pi/(\sqrt{3} a_M)](\pm 1/2, \sqrt{3}/2)$, and $T_j$ with $j=1, 2$ and 3 are hermitian matrices. $\tilde{\mathcal{H}}_{\text{TBG}}$ specified by Eqs.~(\ref{HTBG}) and (\ref{Trr}) has an in-plane periodicity of $a_M$.

We now apply a layer dependent gauge transformation to $\tilde{\mathcal{H}}_{\text{TBG}}$  in order to move the Dirac points at momenta $\boldsymbol{\kappa}_{\pm}$ to the moir\'e Brillouin zone center:
\begin{equation}
\begin{aligned}
&\mathcal{H}_{\text{TBG}} = \begin{pmatrix}
e^{-i \boldsymbol{\kappa}_+ \cdot \rr} & 0\\
0  & e^{-i \boldsymbol{\kappa}_- \cdot \rr}
\end{pmatrix}  
\tilde{\mathcal{H}}_{\text{TBG}}
\begin{pmatrix}
e^{+i \boldsymbol{\kappa}_+ \cdot \rr} & 0\\
0  & e^{+i \boldsymbol{\kappa}_- \cdot \rr}
\end{pmatrix} 
= \begin{pmatrix}
h_0(\kk_\parallel)  & T(\rr) \\
T^{\dagger}(\rr) & h_1(\kk_\parallel)
\end{pmatrix}, \\
&h_0(\kk_\parallel)=\hbar v_F \kk_\parallel \cdot \boldsymbol{\sigma}, \,\,\,\,
h_1(\kk_\parallel)=\hbar v_F  e^{i \theta \sigma_z} (\kk_\parallel \cdot \boldsymbol{\sigma}),\\
& T(\rr) = e^{-i \boldsymbol{\kappa}_+ \cdot \rr} \tilde{T}(\rr) e^{+i \boldsymbol{\kappa}_- \cdot \rr} = \sum_{j=1,2,3} T_j e^{i \bg_{j} \cdot \rr},
\end{aligned}
\label{HHTBG}
\end{equation}
where $\bg_1 = (0, 4\pi/3 a_M)$, and $\bg_{2,3}$ are related to $\bg_1$ by $\pm 2\pi/3$ rotation. After the gauge transformation, the twist relation becomes apparent as $ h_1[\hat{R}(\theta)\kk_\parallel] = h_0(\kk_\parallel) $. We note that $|\bg_1|=|\bb_+|/\sqrt{3}$, and therefore, the new Hamiltonian $\mathcal{H}_{\text{TBG}}$ has an in-plane periodicity of $\tilde{a}_M =\sqrt{3} a_M$.

The new Hamiltonian $\mathcal{H}_{\text{TBG}}$ serves as a 2D building block for chiral twisted graphite, of which the effective Hamiltonian $\mathcal{H}$ is given by
\begin{equation}
\begin{aligned}
&\mathcal{H}(\kk_\parallel, k_z, \rr) = h(\kk_\parallel)+\Delta(k_z,\rr),\\
&h(\kk_\parallel) = \hbar v_F \kk_\parallel \cdot \boldsymbol{\sigma},\\
&\Delta(k_z,\rr)= e^{i k_z} T(\rr) + e^{-i k_z} T^{\dagger}(\rr),
\end{aligned}
\label{HTG}
\end{equation}
where $T(\rr)$ is specified above in Eq.~(\ref{HHTBG}).

We list symmetries of $\mathcal{H}$. By construction,  $\mathcal{H}$ has an in-plane periodicity of $\tilde{a}_M = \sqrt{3} a_M$, and respects both $\hat{C}_{3z}$ and $\hat{C}_{2z} \hat{\mathcal{T}}$ symmetries. In addition, $\mathcal{H}$ has two emergent symmetries described as follows. (1) A translation like symmetry $\mathcal{H}(\kk_{\parallel}, k_z+2\pi/3, \rr+a_M \hat{y})=\mathcal{H}(\kk_{\parallel}, k_z, \rr)$. Due to this symmetry, the energy spectra at $k_z$ and $k_z+2\pi/3$ are identical, and physical quantities such as electron density have a smaller spatial period of $a_M$ compared to $\tilde{a}_M$.  (2) A particle-hole like symmetry $\mathcal{H}(-\kk_{\parallel}, \pi-k_z, -\rr)=-\mathcal{H}(\kk_{\parallel}, k_z, \rr)$. 

The above symmetries can have important implications on Weyl points. (1) The Weyl points at $\kk_{0}=(0,0,0)$ and $\kk_{1}=(0,0,\pi)$ are symmetry enforced due to the $\hat{C}_{3z}$ and $\hat{C}_{2z} \hat{\mathcal{T}}$ symmetries. (2) The particle-hole like symmetry pins the Weyl point at $\kk_{1/2}=(0,0,\pi/2)$ to zero energy.  It is important to note that while some properties (e.g., energy and/or momentum) of the Weyl nodes are influenced by these symmetries,  the very existence of the Weyl nodes are not.

\section{Perturbation theory for the magic angle in chiral twisted graphite}
\label{sec:3}

We present an analytical perturbation theory for the Weyl node at $\kk_{1/2}=(0,0,\pi/2)$  by truncating the corresponding plane-wave expansion of $\mathcal{H}$ to the first momentum shell:
\begin{equation}
\mathcal{H} \approx \begin{pmatrix}
h(\boldsymbol{0})  &i T_1     &i T_2     &i T_3     &-i T_1   &-i T_2   &-i T_3 \\
-i T_1             &h(-\bg_1) &O         &O         &O        &i T_3    &i T_2  \\
-i T_2             &O         &h(-\bg_2) &O         &i T_3    &O        &i T_1  \\
-i T_3             &O         &O         &h(-\bg_3) &i T_2    &i T_1    &O      \\ 
i T_1              &O         &-i T_3    &-i T_2    &h(\bg_1) & O       &O      \\
i T_2              &-i T_3    &O         &-i T_1    &O        &h(\bg_2) &O      \\
i T_3              &-i T_2    &-i T_1    &O         &O        &O        &h(\bg_3)
\end{pmatrix},
\label{HTrunc}
\end{equation}
which are expanded using the 7 in-plane momenta including $\boldsymbol{0}$ and $\pm \bg_{1,2,3}$. In Eq.~(\ref{HTrunc}), $k_z$ has been set to $\pi/2$, and $O$ represents a $2\times 2$ zero matrix. We find that the truncated Hamiltonian in Eq.~(\ref{HTrunc}) has two zero-energy solutions with eigenvectors given respectively by
\begin{equation}
\begin{aligned}
&|\Phi_A\rangle =\frac{1}{\mathcal{N}_A}\{(1,0),\psi_1,\psi_2,\psi_3,\psi_1,\psi_2,\psi_3\}^{T},\\
&|\Phi_B\rangle =\frac{1}{\mathcal{N}_B}\{(0,1),\varphi_1,\varphi_2,\varphi_3,\varphi_1,\varphi_2,\varphi_3\}^{T},
\end{aligned}
\end{equation}
where $\mathcal{N}_{A, B}$ are normalization factors. The two-component spinors $\psi_{j}$ and $\varphi_{j}$ are found to be:
\begin{equation}
\begin{aligned}
&\psi_{j} = \Big(\frac{\alpha_1^2-\alpha_1-\alpha_0^2}{1-2\alpha_0^2-\alpha_1^2}, \frac{\alpha_0 (1+3 \alpha_1)}{1-2\alpha_0^2-\alpha_1^2}e^{i (j-1) 2\pi/3}  \Big)   ,\\
&\varphi_{j} = \Big( \frac{\alpha_0(3\alpha_1-1)}{1-2\alpha_0^2-\alpha_1^2} e^{-i (j-1) 2\pi/3}, \frac{\alpha_1^2+\alpha_1-\alpha_0^2}{1-2\alpha_0^2-\alpha_1^2} \Big),
\end{aligned}
\end{equation}
where $(\alpha_0,\alpha_1)=(w_{AA},w_{AB})/(\hbar v_F |\bg_1|)$. Here $\psi_{1, 2, 3}$ are related by $\pm 2\pi/3$ rotations in the spinor space, and the same is also true for $\varphi_{1, 2, 3}$. By projecting the Hamiltonian $\mathcal{H}$ onto the subspace of $\{|\Phi_A\rangle, |\Phi_B\rangle \}$  to linear order in $\kk_{\parallel}$ and $q_z=k_z-\pi/2$, we obtain an effective Hamiltonian for the Weyl fermion at $\kk_{1/2}$
\begin{equation}
\mathcal{H}_{\text{Weyl}}=\hbar \begin{pmatrix}
v_1 q_z & v_{\parallel} (k_x-i k_y) \\
v_{\parallel} (k_x+i k_y)  & v_2 q_z
\end{pmatrix},
\label{HWeylG}
\end{equation}
where the in-plane velocity $v_{\parallel}$ from this perturbation theory is given by
\begin{equation}
v_{\parallel}/v_F =\frac{1}{\mathcal{N_A} \mathcal{N_B}} \frac{1-4 \alpha _0^2+10 \alpha _0^4-8 (1+\alpha _0^2) \alpha _1^2+7 \alpha _1^4}{(1-2\alpha_0^2-\alpha_1^2)^2}.
\label{vpara}
\end{equation}
The analytical expression in Eq.~(\ref{vpara}) indicates that $v_{\parallel}$  vanishes at two magic angles $\theta_{M,1}^* \approx 1.87^\circ$ and $\theta_{M,2}^* \approx 0.46^\circ$, which are obtained using numerical values of $(v_F, w_{AA}, w_{AB})$ listed in the main text. By comparison, the magic angles given by the full  band structure calculations are $\theta_{M,1} \approx 1.67^{\circ}$ and $\theta_{M,2} \approx 1.09^{\circ}$. Thus, $\theta_{M,1}^*$ provides a semiquantitative estimation of $\theta_{M,1}$, but $\theta_{M,2}^*$ differs significantly from $\theta_{M,2}$, as the truncated Hamiltonian in Eq.~(\ref{HTrunc}) becomes less accurate at smaller twist angles. We have also checked that the out-of-plane velocities $v_1$ and $v_2$ are generally different within the perturbation theory, confirming that the effective Hamiltonian $\mathcal{H}_{\text{Weyl}}$ expanded around $\kk_{1/2}$ indeed describes Weyl fermion.

\section{Interlayer tunneling matrix in twisted Weyl semimetal}
\label{sec:4}
In our model of twisted Weyl semimetal, each layer has a square lattice structure, and each site hosts two orbitals $|j_z=
\frac{3}{2}\rangle$ and $|j_z=\frac{1}{2}\rangle$. To derive the interlayer tunneling between two twisted layers, we use a two-step approach under a local approximation.

In the first step, we consider two layers with a zero twist angle ($\theta=0$) but with a finite in-plane displacement $\dd$. The $\dd$-dependence of the interlayer tunneling is denoted as $w_{\lambda_1 \lambda_2}(\dd)  = \langle \lambda_1 | H_T | \lambda_2 \rangle_{\dd}$, where $\lambda_{1,2}$ label orbitals that reside  respectively in layer $1$ and $2$, $| \lambda_i \rangle$ represents the corresponding Bloch wave state with a zero in-plane crystal momentum, and $H_T$ is the interlayer tunneling Hamiltonian. Using the real space representation of Bloch waves, we can express $w_{\lambda_1 \lambda_2}(\dd)$ as
\begin{equation}
\begin{aligned}
w_{\lambda_1 \lambda_2}(\dd) &  = \frac{1}{N_\parallel} \sum_{\RR_1, \RR_2} \langle \lambda_1, \RR_1 | H_T | \lambda_2, \RR_2+\dd \rangle,
\end{aligned}
\end{equation}
where $N_\parallel$ is the number of unit cells in each layer, and $\RR_{1,2}$ indicate lattice sites in the square lattice of layer 1. We now make a two-center approximation:
\begin{equation}
t_{\lambda_1 \lambda_2}(\boldsymbol{\delta}) = \langle \lambda_1, \RR | H_T | \lambda_2, \RR+\boldsymbol{\delta} \rangle,
\label{tll}
\end{equation}
which is the hopping integral between an orbital $\lambda_1$ at position $\RR$ and another orbital $\lambda_2$ at position $\RR+\boldsymbol{\delta}$, and is assumed to be independent of the position $\RR$. We further perform a Fourier transformation: $t_{\lambda_1 \lambda_2}(\boldsymbol{\delta} )= (1/N_\parallel) \sum_{\qq} t_{\lambda_1 \lambda_2}(\qq) e^{i \qq \cdot \boldsymbol{\delta} } $. Then $w_{\lambda_1 \lambda_2}(\dd)$ can be simplified to be 
\begin{equation}
\begin{aligned}
w_{\lambda_1 \lambda_2}(\dd) & = \frac{1}{N_\parallel}  \sum_{\RR_1, \RR_2} t_{\lambda_1 \lambda_2} (\RR_2+\dd-\RR_1) \\
& = \frac{1}{N_\parallel^2} \sum_{\RR_1, \RR_2} \sum_{\qq} e^{i \qq \cdot (\RR_2+\dd-\RR_1)} t_{\lambda_1 \lambda_2} (\qq)\\
&= \sum_{\GG} t_{\lambda_1 \lambda_2}(\GG) e^{i \GG \cdot \dd}  ,
\end{aligned}
\label{wt}
\end{equation}
where $\GG$ is the reciprocal lattice vector of the square lattice. The final expression in Eq.~(\ref{wt}) reflects the fact that $w_{\lambda_1 \lambda_2}(\dd)$ is a periodic function of $\dd$ with the square lattice periodicity.

In the second step, we now consider two twisted layers with a finite small twist angle ($\theta \neq 0$). In a long-period moir\'e pattern, the local in-plane displacement $\dd$ between the two layers has a spatial variation approximated by $\theta \hat{z} \times \rr$, where $\rr$ represents the coarse-grained in-plane position. We make a local approximation, assuming that the interlayer tunneling varies in space following the local displacement $\dd (\rr)$. Replacing $\dd$ in Eq.~(\ref{wt}) by $\theta \hat{z} \times \rr$, we obtain interlayer tunneling terms for the two twisted layers
\begin{equation}
w_{\lambda_1 \lambda_2}(\rr) = \sum_{\GG} t_{\lambda_1 \lambda_2}(\GG) e^{i \bg \cdot \rr},
\end{equation}
where $\bg= \theta \GG \times \hat{z}$ represents the moir\'e reciprocal lattice vector, and $w_{\lambda_1 \lambda_2}(\rr)$ has the moir\'e periodicity of $a_M = a_0/\theta$.

\begin{figure}[t]
	\includegraphics[width=0.8\columnwidth]{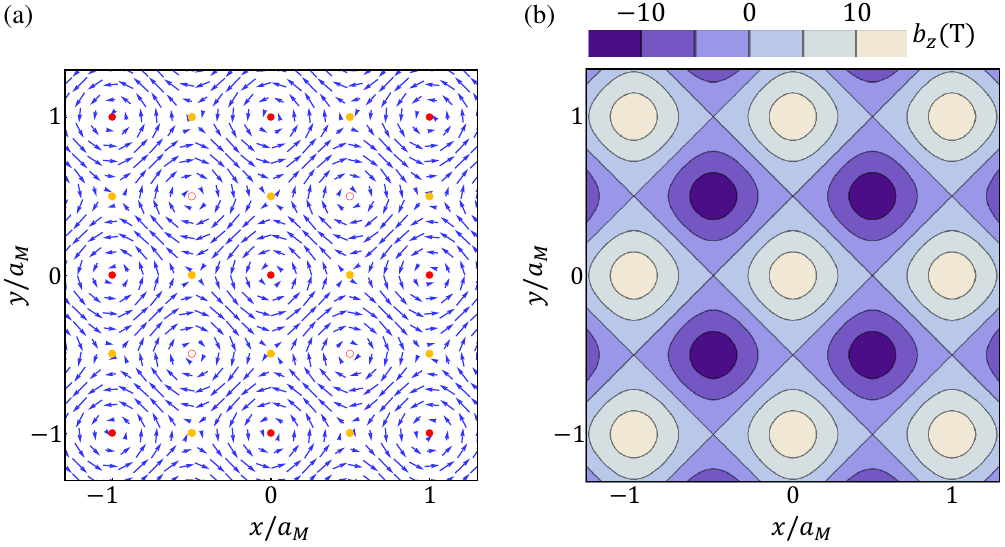}
	\caption{(a) 2D plot of $\boldsymbol{\mathcal{A}}$ field (represented by arrows). $t_{sp}/v_W>0$ and $k_z=\pi/2$ have been assumed. Both the red dots ($\boldsymbol{\mathcal{R}}$ positions) and circles ($\boldsymbol{\mathcal{R}}_{1/2}$ positions) indicate positions of vortex cores of the $\boldsymbol{\mathcal{A}}$ field, but the corresponding pseudo magnetic field $b_z$ is respectively  positive and negative at these two types of positions. The yellow dots are positions of antivoretx cores, where the corresponding $b_z$ field vanishes. (b) Corresponding 2D map of the $b_z$ field. }
	\label{Fig4}
\end{figure}

We now use rotational symmetry to put further constraints on the tunneling terms. For intraorbital tunneling ($\lambda_1=\lambda_2$), $t_{\lambda \lambda}(\boldsymbol{\delta})$ in Eq.~(\ref{tll}) equals $t_{\lambda \lambda}[\hat{R}(\frac{\pi}{2})\boldsymbol{\delta}]$ because of the rotational symmetry, and similarly, in the reciprocal space, $t_{\lambda \lambda}(\GG) = t_{\lambda \lambda}[\hat{R}(\frac{\pi}{2})\GG]$. We expect that $t_{\lambda \lambda}(\qq)$ in the reciprocal space should decay rapidly for $|\qq| d_z >1$ \onlinecite{Bistritzer2011}, where $d_z$ is the interlayer distance. Therefore, we can  just keep the $\bg_0=\GG_0=\boldsymbol{0}$ term in the intraorbital tunneling $w_{\lambda \lambda}(\rr)$, which then has no spatial variation under this approximation.

The rotational symmetry plays a different role for interorbital tunneling. For definiteness, we take $\lambda_1=\frac{1}{2}$ and $\lambda_2=\frac{3}{2}$ to respectively represent $|j_z=\frac{1}{2}\rangle$ and $|j_z=\frac{3}{2}\rangle$ states. Because the two states have different angular momenta, $t_{\frac{1}{2}, \frac{3}{2}}[\hat{R}(\frac{\pi}{2})\boldsymbol{\delta}] = \exp(i \frac{\pi}{2}) t_{\frac{1}{2}, \frac{3}{2}}(\boldsymbol{\delta})$. Correspondingly, in reciprocal space, $t_{\frac{1}{2}, \frac{3}{2}}[\hat{R}(\frac{\pi}{2})\GG] = i t_{\frac{1}{2}, \frac{3}{2}}(\GG)$, which implies that $t_{\frac{1}{2}, \frac{3}{2}}(\GG_0=\boldsymbol{0})$ vanishes. Thus, the lowest moir\'e harmonics that contribute to $w_{\frac{1}{2}, \frac{3}{2}}(\rr)$ are from the first shell of moir\'e reciprocal lattice vectors with $|\bg|=2\pi/a_M$. As a result, the interlayer tunneling matrix in the basis of  $|j_z=\frac{3}{2}\rangle$ and $|j_z=\frac{1}{2}\rangle$  can be parametrized as
\begin{equation}
T(\rr)= \begin{pmatrix}
t_{pp} & 0 \\
0 & t_{ss}
\end{pmatrix}+
t_{sp}\sum_{\bg}^{'} e^{i \bg \cdot \rr}
\begin{pmatrix}
0 &   i e^{-i \phi_{\bg}} \\
- i e^{i \phi_{\bg}} & 0
\end{pmatrix},
\end{equation}
where $t_{ss}$, $t_{pp}$ and $t_{sp}$ are tunneling parameters, and the summation over $\bg$ is restricted to the first shell. The total effective Hamiltonian for the twisted Weyl semimetal is
\begin{equation}
\begin{aligned}
\mathcal{H}_W&= \hbar v_W \kk_{\parallel} \cdot \boldsymbol{\sigma} - (M_0\cos Q_z+M_1 \kk_{\parallel}^2) \sigma_z + e^{i k_z} T(\rr) + e^{-i k_z} T^{\dagger}(\rr) \\
&=\begin{pmatrix}
M(\kk) & \hbar v_W (k_x - i k_y)  \\
\hbar v_W (k_x + i k_y) & -M(\kk)
\end{pmatrix}
+ 2 {t_{sp}} \sin k_z \sum_{\bg}^{'} \begin{pmatrix}
0 &  e^{-i (\phi_{\bg}+\bg\cdot \rr)} \\
e^{i (\phi_{\bg}+\bg\cdot \rr)} & 0
\end{pmatrix},
\end{aligned}
\label{HWW}
\end{equation}
where $M(\kk)=M_0(\cos k_z-\cos Q_z)-M_1 \kk_{\parallel}^2$, and $M_0=2t_{pp}=-2t_{ss}$. We have chosen the parameter $t_{sp}$ to be real such that it is consistent with the intralayer term $\hbar v_W \kk_{\parallel} \cdot \boldsymbol{\sigma}$. Based on Eq.~(\ref{HWW}), we can define an effective gauge field $\boldsymbol{\mathcal{A}}$ as
\begin{equation}
\boldsymbol{\mathcal{A}} = \frac{2 {t_{sp}}}{e v_W} \sin k_z \sum_{\bg}' \Big( \cos(\bg\cdot \rr +\phi_{\bg} ),\,\,\, \sin(\bg\cdot \rr +\phi_{\bg} ) \Big).
\end{equation}
Because $\boldsymbol{\mathcal{A}}$ is proportional to $\sin k_z$, it represents a chiral gauge field that couples oppositely to Weyl nodes with different topological charges. An illustration of $\boldsymbol{\mathcal{A}}$ field in the moir\'e pattern is demonstrated in Fig.~\ref{Fig4}, which shows that the $\boldsymbol{\mathcal{A}}$ field has a vortex-antivortex lattice configuration.

\end{widetext}

\end{document}